\documentclass[%
 reprint,
 superscriptaddress,
 amsmath,amssymb,
 aps,
]{revtex4-2}

\usepackage{graphicx}
\usepackage{dcolumn}
\usepackage{bm}
\usepackage{physics}

\begin{document}

\title{One-hour coherent optical storage in an atomic frequency comb memory}

\author{Yu Ma}
\author{You-Zhi Ma}
\author{Zong-Quan Zhou}
 \email{email: zq\_zhou@ustc.edu.cn; cfli@ustc.edu.cn}
\author{Chuan-Feng Li}
 \email{email: zq\_zhou@ustc.edu.cn; cfli@ustc.edu.cn}
\author{Guang-Can Guo}
\affiliation{
 CAS Key Laboratory of Quantum Information,\\University of Science and Technology of China, Hefei, 230026, P. R. China
}
\affiliation{
 CAS Center For Excellence in Quantum Information and Quantum Physics, University of Science and Technology of China, Hefei, 230026, P. R. China
}

\date{\today}

\maketitle

\section*{abstract}

Photon loss in optical fibers prevents long-distance distribution of quantum information on the ground. Quantum repeater is proposed to overcome this problem, but the communication distance is still limited so far because of the system complexity of the quantum repeater scheme. Alternative solutions include transportable quantum memory and quantum-memory-equipped satellites, where long-lived optical quantum memories are the key components to realize global quantum communication. However, the longest storage time of the optical memories demonstrated so far is approximately 1 minute. Here, by employing a zero-first-order-Zeeman magnetic field and dynamical decoupling to protect the spin coherence in a solid, we demonstrate coherent storage of light in an atomic frequency comb memory over 1 hour, leading to a promising future for large-scale quantum communication based on long-lived solid-state quantum memories.

\clearpage

\section*{Introduction}

Quantum repeater \cite{repeater_1998} is proposed to establish the long-distance entanglement in the presence of channel loss. However, the construction of a practical quantum repeater is of great challenge due to its demanding requirements and system complexity \cite{review_repeater,network_duan,single-atom_qm,entanglement_nv,memory_entanglement}. Space-borne quantum communication \cite{space-borne,satellite_entang} and transportable quantum memory \cite{6hour} can avoid using optical fibers, but optical quantum memories with a lifetime on the order of hours would be essential for extending the communication distance to global scale.

With the help of a zero-first-order-Zeeman (ZEFOZ) magnetic field and dynamical decoupling (DD), Zhong et al. have reported a six-hour spin coherence lifetime \cite{6hour} in europium-doped yttrium orthosilicate (Eu$^{3+}$:Y$_2$SiO$_5$). This long-lived spin coherence is a key resource for constructing quantum memories for global quantum communication. Although hours of spin coherence time have been demonstrated, long-lived optical storage remains a challenge because of the complicated and unknown energy structures in ZEFOZ fields \cite{longlifeqm,6hour,my_2018} and a reduced effective absorption in magnetic fields \cite{euyso151-swafc-dd} because only one subsite can be used for the long-term storage \cite{my_2018}. To date, the longest optical storage time is approximately one minute realized in $^{87}$Rb atoms \cite{rb87-1min} and a Pr$^{3+}$:Y$_2$SiO$_5$ crystal using the electromagnetically induced transparency protocol \cite{longlifeqm}. For single-photon-level storage, the longest storage time is approximately one second realized in $^{87}$Rb atoms \cite{Rb_0.22s}.

To take advantages of the long-lived spin coherence, a spin-wave based optical storage protocol should be employed. So far, atomic frequency comb (AFC) \cite{afcexp_1st,afc_theo} has been the only successful protocol for spin-wave storage of photonic qubits \cite{pryso-sw-qm,time_entang,multiplexed_yang} in the rare-earth-ion doped crystals. In Eu$^{3+}$:Y$_2$SiO$_5$, spin-wave storage of polarization qubit \cite{EuYSO_multiplexed} and the generation and multimode storage of single photons as spin waves \cite{euyso-correlation} have been demonstrated based on AFC protocol. A recent work extends the AFC memory time to 0.5 seconds in Eu$^{3+}$:Y$_2$SiO$_5$ \cite{euyso-0.53s}.

Here we demonstrate a coherent optical memory with a storage time of one hour using a $^{151}$Eu$^{3+}$:Y$_2$SiO$_5$ crystal, based on the spin-wave AFC protocol \cite{afc_theo} in a ZEFOZ field, namely ZEFOZ-AFC method. Dynamical decoupling is used to protect the spin coherence and extend the storage time. The coherent nature of this device is verified by implementing a time-bin-like interference experiment after 1-hour storage with a fidelity of 96.4\%, which shows the feasibility of qubit storage. Our work improves the AFC memory time by approximately 6000 times \cite{euyso-0.53s}.

\begin{figure*}[t]
\includegraphics[width=\textwidth]{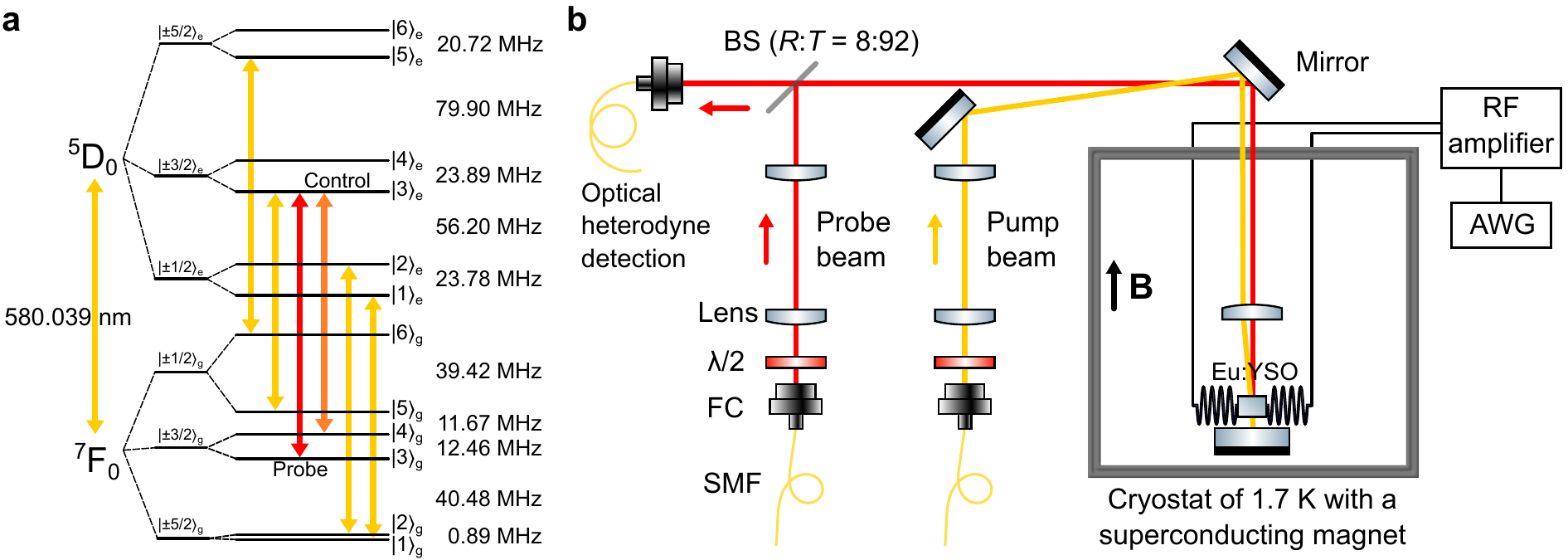}
\caption{\label{fig:setup}\textbf{Energy level diagram and experimental setup.} \textbf{a} The laser at 580.039 nm is resonant with electronic state $^5$D$_0$ and $^7$F$_0$ of $^{151}$Eu$^{3+}$:Y$_2$SiO$_5$, which split into six hyperfine levels respectively in the ZEFOZ magnetic field. $\ket{3}_\mathrm{g}$, $\ket{4}_\mathrm{g}$ and $\ket{3}_\mathrm{e}$ form the $\Lambda$ system for the long-lived spin-wave AFC storage. Details about the hyperfine structure and the pumping scheme can be found in the Supplementary Note 2. The AFC is prepared in the $\ket{3}_\mathrm{g}$ level. The control pulse resonant with the $\ket{4}_\mathrm{g}$ and $\ket{3}_\mathrm{e}$ transfers the optical coherence into a spin wave and drives the spin wave back to the optical regime. \textbf{b} Schematic of the experimental setup. The probe and pump beam are emitted with single-mode fibers (SMF) and fiber collimators (FC) before entering the cryostat. The two beams are arranged in a non-collinear configuration. Half-wave plates ($\lambda$/2) control the polarization of the beams, and two pairs of lenses are used to obtain proper beam widths at the position of the crystal. A BS (beam splitter) with a reflection-to-transmission ratio of 8:92 is employed to efficiently collect the transmitted signal. The probe beam is reflected by the mirror at the bottom of the cryostat and coupled to a SMF for optical heterodyne detection after passing through the BS. A pair of coils placed at the two sides of the sample is driven by an arbitrary waveform generator (AWG) with the output amplified with a 300-W RF amplifier.}
\end{figure*}

\section*{Results}

\textbf{Characterization of the energy level structure.} In order to implement optical storage in a ZEFOZ field, the knowledge of the energy level structures in both the ground and excited state is a prerequisite. Previous works \cite{euyso-ground_rhs,euysoexcitedgisin,my_2018} have determined the spin Hamiltonians for the ground state $^7$F$_0$ and excited state $^5$D$_0$, which can be used to predict the level structures in any given magnetic field. However, the theoretical predictions may have an error comparable to the storage bandwidth and lead to a wrong choice of pumping strategies. In order to precisely determine the structures, we first find the ZEFOZ transition $\ket{3}_\mathrm{g}\leftrightarrow\ket{4}_\mathrm{g}$ by adjusting the strength of the field and orientation of the sample. We then use the continuous-wave Raman heterodyne detection (RHD) \cite{ramanprl} to probe the ground-state resonances in the ZEFOZ field. However, continuous-wave RHD fails to detect the excited-state resonances due to the short-lived population in the optically excited states and the weaker interaction strength with the radio-frequency (RF) field. Here we employ the pulsed RHD to probe the excited-state resonances (see Supplementary Note 1). The experimentally determined energy level structure is presented in Fig.~\ref{fig:setup}a.

\textbf{Experimental setup.} In order to increase the effective absorption, we use an isotopically enriched sample of $^{151}$Eu$^{3+}$:Y$_2$SiO$_5$ crystal (Scientific Materials), with a doping level of 1000 ppm and an isotopic enrichment of 99.3\%. The sample has an optical depth $\alpha L$ of 2.6 for Eu$^{3+}$ at site 1. In order to quickly find the correct direction of the sample in the ZEFOZ field, the optical surface of the sample is cut perpendicularly to the direction of the known ZEFOZ magnetic field with a cylindrical shape, such that the field of the superconductor magnet along the vertical direction (Fig.~\ref{fig:setup}b) is approximately aligned with the ZEFOZ field (a photo is provided in Supplementary Fig.~2b). The field is 1.280 T in the direction of $[-0.535, -0.634, 0.558]$ in the $[D_1, D_2, b]$ frame of the crystal as indicated in previous works \cite{6hour,my_2018}). The sample has a diameter of 4.5 mm and a thickness of 6 mm, and is mounted on two goniometers. The goniometers allow tilting in two dimensions, with a tilting range within $6.6^\circ$ and a positioning resolution of $0.002^\circ$. A pair of 4-turn Helmholtz coils with a diameter of 6 mm is placed at the two sides of the sample. The coils are driven by an arbitrary waveform generator (AWG) with the output amplified by an RF amplifier with an output power of 300 W, to implement the DD sequences. The experiments are conducted after the sample is cooled to 1.7 K.

The laser is locked to an ultra-stable plano-concave cavity at 580.039 nm, with a linewidth well below 10 kHz. The pump and probe beam are independently generated by acoustic-optic modulators (AOM) in the double-pass configuration. Then the pump and probe beam are overlapped with each other inside the crystal with a diameter of 200 and 50 $\mu$m respectively, and an angle of approximately 30 mrad with each other. The beams double pass the crystal, and the transmitted probe beam is collected by a single-mode fiber. The probe beam is then combined with a reference laser at a locked frequency offset of 43.66 MHz for optical heterodyne detection. The beat signal is captured by a photodetector and demodulated to give the amplitude of the signals.

\begin{figure*}[t]
\includegraphics[width=\textwidth]{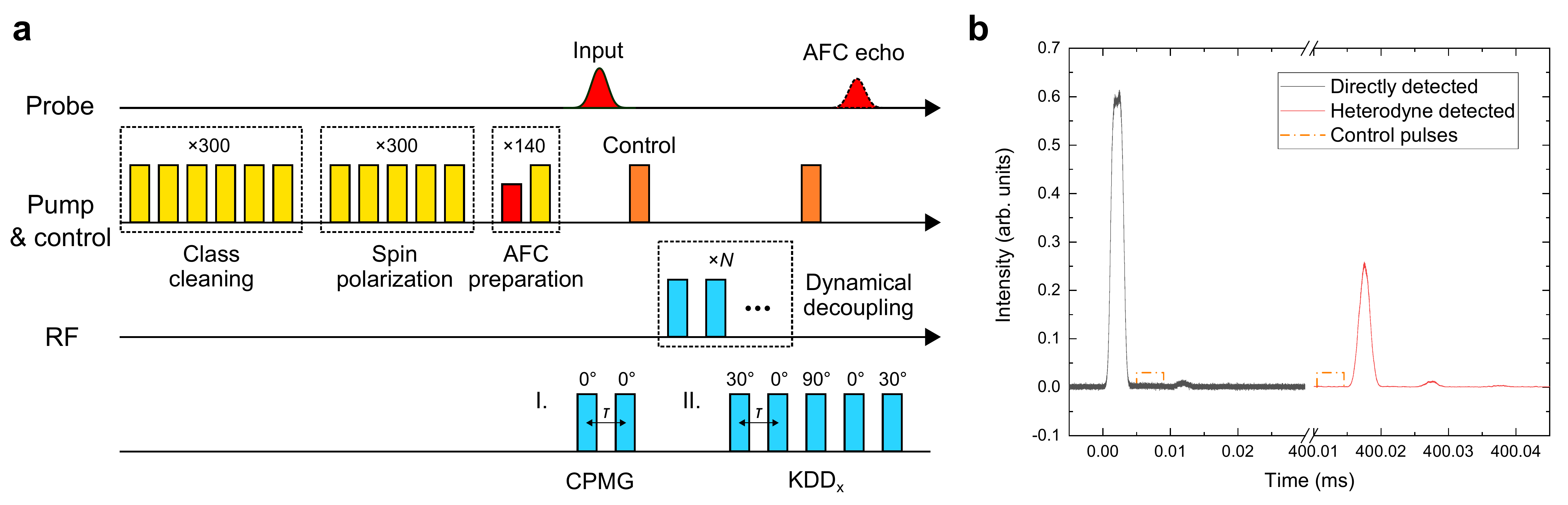}
\caption{\label{fig:sequence}\textbf{Experimental sequence and oscilloscope traces.} \textbf{a} The experimental sequence. The class cleaning sequence containing six pump lights is used to pump away unwanted ions and select a single class of ions. The spin polarization sequence polarizes the population into the $\ket{3}_\mathrm{g}$ state. Then the AFC is prepared in the $\ket{3}_\mathrm{g}$ state while keeping the $\ket{4}_\mathrm{g}$ empty using the $\ket{4}_\mathrm{g}\leftrightarrow\ket{3}_\mathrm{e}$ pump light. After the input pulse (Gaussian profile, colored with red) is absorbed, a control pulse transfers the optical excitation into a spin-wave excitation for long-term storage. Dynamical decoupling is employed for protecting the spin coherence and finally the spin excitation is transferred back to the optical regime with another control pulse. Two kinds of dynamical decoupling sequences, CPMG and KDD$_\mathrm{x}$, are tested for spin coherence protection. \textbf{b} Oscilloscope traces of the spin-wave AFC storage with a CPMG sequence with $\tau = 100$ ms, which consists of four $\pi$ pulses. The black line shows the traces of transmission of the input mode and remaining two-level AFC echo, which are directly detected by the photodetector. The red line shows the trace of the spin-wave AFC echo after a storage of 400 ms, which is heterodyne detected. The traces are averaged four times. The dashed line indicates the position of the control pulses.}
\end{figure*}

\begin{figure*}[t]
\includegraphics[width=\textwidth]{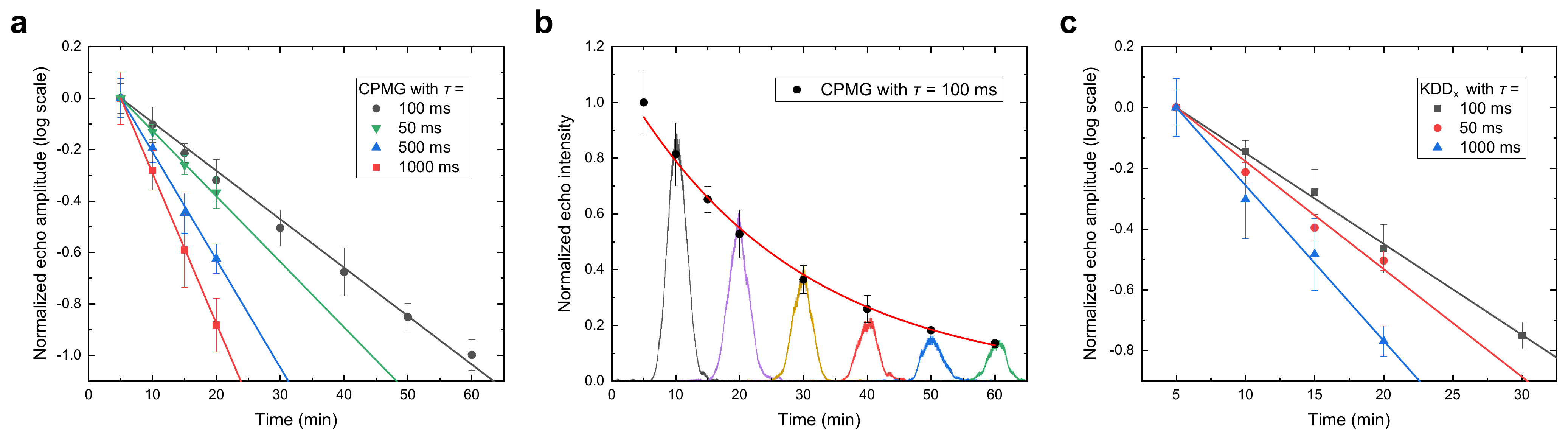}
\caption{\label{fig:dd}\textbf{Measurements of the memory lifetime.} \textbf{a, c} Decay of the spin-wave AFC echo with the CPMG and KDD$_\mathrm{x}$ sequences, respectively. The two DD sequences reach their longest storage lifetime of $52.9\pm1.2$ and $33.3\pm1.1$ minutes with a period $\tau = 100$ ms. \textbf{b} The decay of the echo intensity using CPMG with $\tau = 100$ ms versus the storage time. Some of the single-shot oscilloscope traces of the readout signals are included. Each data point is averaged four times. The error bars indicate one standard deviation of the echo area.}
\end{figure*}

\textbf{Spin-wave AFC and dynamical decoupling.} The experimental sequence is shown in Fig.~\ref{fig:sequence}a. The experiment begins with a preparation procedure based on spectral hole burning \cite{SHBEuYSO} (see Supplementary Note 2). The first step is the so-called ``class-cleaning", which can isolate a single class of ions in the inhomogeneously broadened sample. Six chirped pulses which are resonant with $\ket{1}_\mathrm{g}\leftrightarrow\ket{1}_\mathrm{e}$, $\ket{2}_\mathrm{g}\leftrightarrow\ket{2}_\mathrm{e}$, $\ket{3}_\mathrm{g}\leftrightarrow\ket{3}_\mathrm{e}$, $\ket{4}_\mathrm{g}\leftrightarrow\ket{3}_\mathrm{e}$, $\ket{5}_\mathrm{g}\leftrightarrow\ket{3}_\mathrm{e}$ and $\ket{6}_\mathrm{g}\leftrightarrow\ket{5}_\mathrm{e}$ transitions are applied to choose the target ions as shown in Fig.~\ref{fig:setup}a. Other classes of ions will relax to other non-resonant ground-state levels and no longer interact with the six pulses. The chirp bandwidth of these pulses is 1 MHz. The second step is the so-called ``spin polarization", which initializes all the population of the chosen ions into the state $\ket{3}_\mathrm{g}$ by using the five pulses mentioned above except for $\ket{3}_\mathrm{g}\leftrightarrow\ket{3}_\mathrm{e}$. After these two steps, a $\Lambda$ system has been well isolated, which comprises the level $\ket{3}_\mathrm{g}$, $\ket{4}_\mathrm{g}$ and $\ket{3}_\mathrm{e}$. The third step is the ``AFC preparation". An AFC is prepared in the $\ket{3}_\mathrm{g}$ level by frequency-selected hole burning while keeping the $\ket{4}_\mathrm{g}$ level empty. The AFC has a bandwidth of 1.0 MHz and a periodicity of $\Delta = 100$ kHz, which leads to an echo emission at $t = 1/\Delta = 10$ $\mu$s after input. After the AFC preparation, a 2-$\mu$s (full width at half maximum) probe pulse of Gaussian profile with a power of 150 $\mu$W resonant with $\ket{3}_\mathrm{g}\leftrightarrow\ket{3}_\mathrm{e}$ is absorbed by the sample and creates a collective excitation. A control pulse with a width of 4 $\mu$s and a power of 360 mW resonant with $\ket{4}_\mathrm{g}\leftrightarrow\ket{3}_\mathrm{e}$ immediately transfers the optical coherence into a spin-wave excitation. The control pulse has a complex hyperbolic secant profile \cite{CHS} to achieve efficient control over the bandwidth of the AFC. Then a DD sequence is applied to protect the spin coherence. After that, a second control pulse transfers the spin coherence back to the optical excitation. The collective state will continue rephasing and emits an AFC echo when the excited-state storage time reaches $1/\Delta = 10$ $\mu$s.

\begin{figure*}[htb]
\includegraphics[width=\textwidth]{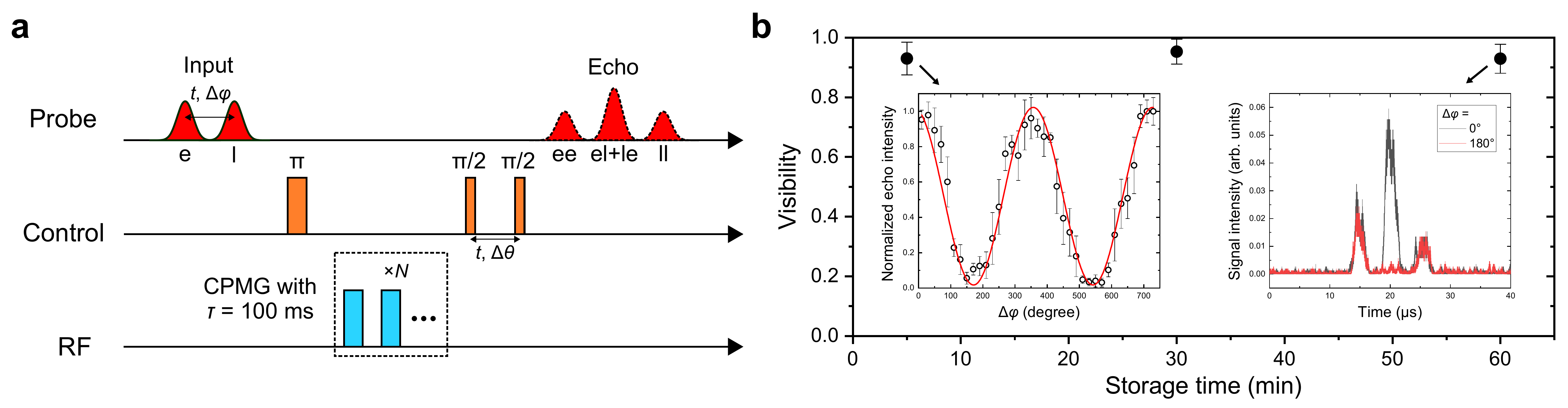}
\caption{\label{fig:interf}\textbf{Verification of the coherent nature of the memory.} \textbf{a} Experimental sequence of the AFC echo interference, where e and l represent the earlier and the later, respectively. The interval of the two input pulses and that of the two readout pulses ($\pi$/2) should be the same in order to overlap the echoes (el+le) of the two inputs. \textbf{b} Visibility versus the storage time. The left inset shows an interference pattern after 5-minute storage, generated by recording the intensity of the echo in the middle while varying the relative phase $\Delta\varphi$ between the two input pulses. Each data point is averaged four times, and the error bars indicate one standard deviation of the echo area. After a storage time of 5, 30 and 60 minutes, the coherence is well maintained with a visibility of $93.0\pm5.5\%$, $95.3\pm4.2\%$ and $92.9\pm4.9\%$, respectively. Single-shot oscilloscope traces of the signals after the 60-minute storage are presented in the right inset. Constructive and destructive interference are observed at $\Delta\varphi = 0^{\circ}$ and $180^{\circ}$, respectively.}
\end{figure*}

Two kinds of DD sequences, i.e. CPMG and KDD$_\mathrm{x}$, are used to protect the spin coherence in our experiments. As shown in Fig.~\ref{fig:sequence}a, each of them consists of sequences of $\pi$ pulses with different phase shifts, where $\tau$ is the spacing between the center of the $\pi$ pulses. As a comparison with the optical storage times, the spin coherence times with DD are also measured using Raman heterodyne detection (see Supplementary Note 3). CPMG and KDD$_\mathrm{x}$ can reach a $1/e$ spin coherence lifetime of $2.68\pm0.06$ and $0.95\pm0.03$ hours when $\tau = 20$ ms, respectively. For storage of light, we measure the decay of the echo with various intervals $\tau$ shown in Fig.~\ref{fig:dd}a and \ref{fig:dd}c. The longest storage times are achieved at $\tau = 100$ ms, where CPMG and KDD$_\mathrm{x}$ have a lifetime of $52.9\pm1.2$ and $33.3\pm1.1$ minutes, respectively. These results are in good agreement with the spin coherence times measured with the same $\tau = 100$ ms, which are $50.6\pm2.0$ and $38.2\pm2.0$ minutes (see Supplementary Note 3).

In Fig.~\ref{fig:sequence}b, we present oscilloscope traces of the spin-wave AFC storage with a CPMG sequence with $\tau = 100$ ms, which consists of four $\pi$ pulses. Traces recorded after longer storage times can be found in Fig.~\ref{fig:dd}b. When $\tau$ is shorter than 100 ms, unlike the spin coherence times, the optical storage lifetimes are shortened as shown in Fig.~\ref{fig:dd}a and \ref{fig:dd}c. This is because DD sequences with smaller intervals result in a greater heating effect of the RF coils, which broadens the optical homogeneous linewidth \cite{euyso_spectra-t1-23d}, leading to a reduction of the AFC efficiency.

The storage efficiency is analyzed as follows:
\begin{equation}
\eta_{\mathrm{total}} = \eta_{\mathrm{AFC}}\cdot(\eta_{\mathrm{control}})^{2}\cdot\eta_{\mathrm{spin}}.\label{eqn:efficiency}
\end{equation}
The two-level AFC efficiency $\eta_{\mathrm{AFC}}$ without spin-wave storage and DD is 4.5\%. This efficiency can be enhanced close to unity by using cavity enhancement \cite{pryso_cvty} technique to achieve an efficient photon absorption. We also note that this efficiency will be reduced if a RF sequence is applied to the sample due to the heating. For example, $\eta_{\mathrm{AFC}}$ is reduced to 2.5\% after a DD sequence with $\tau = 100$ ms. The transfer efficiency of the control pulse $\eta_{\mathrm{control}}$ is determined to be 38.5\% (see Supplementary Note 2). This transfer efficiency is not very high because the polarization of the incident light cannot be parallel to D$_1$ axis (maximum absorption) of the crystal due to the special sample orientation described above. The total storage efficiencies for 5-minute storage with CPMG and KDD$_\mathrm{x}$ when $\tau = 100$ ms are 0.035\% and 0.052\%, respectively. $\eta_{\mathrm{spin}}$ is the efficiency of storage in the spin states, which can be estimated from equation (\ref{eqn:efficiency}) to be 9.5\% and 14.1\% for 5-minute storage with CPMG and KDD$_\mathrm{x}$ when $\tau = 100$ ms, respectively. $\eta_{\mathrm{spin}}$ is primarily limited by the insufficient bandwidth of the $\pi$ pulses and the inhomogeneity of the RF coils in our case. The width of the $\pi$ pulse is determined at 65.1 $\mu$s (see Supplementary Note 3). However, the inhomogeneous broadening $\Gamma_{\mathrm{inh}}$ of the spin transition $\ket{3}_\mathrm{g}\leftrightarrow\ket{4}_\mathrm{g}$ of the Eu$^{3+}$ ions in the crystal is 30 kHz, which cannot be entirely covered by the bandwidth of the $\pi$ pulse $\Gamma_{\pi}\approx1/(65.1~\mathrm{\mu s})\approx15$ kHz. The decoherence of those ions, with transition frequencies outside the bandwidth of the $\pi$ pulse, cannot be recovered by the DD sequences, thus leading to an efficiency loss. $\eta_{\mathrm{spin}}$ is also related to the inhomogeneity of the RF field generated by the coils, because the thickness of the sample is comparable to the diameter of the coils. A more efficient spin manipulation can be achieved by RF coils with an improved homogeneity, thus increasing the storage efficiency. The total storage efficiencies can be greatly enhanced by the combination of those techniques mentioned above.

\textbf{Coherence verification.} Quantum information can be encoded into various degrees of freedom of photons, such as frequency, time and polarization. The time-bin encoding has the particular advantage of good robustness to environmental fluctuations. Therefore, here we implement the coherent light storage with time-bin-like encoding \cite{pryso-sw-qm}. The experimental sequence is shown in Fig.~\ref{fig:interf}a. Here the AFC preparation and DD sequences are the same as that described above. In order to verify its long-term coherence protection, CPMG with an interval $\tau=100$ ms is used to extend the spin coherence time. Two input pulses with an interval $t$ are stored in the memory and partially read out with two $\pi/2$ pulses separated by the same $t$, respectively, after the CPMG sequence. Each input pulse is read out as two AFC echoes, and therefore the later echo of the earlier input and the earlier of the later are overlapped. The relative phase $\Delta\varphi$ between the two input pulses is varied to generate the interference pattern presented in Fig.~\ref{fig:interf}b while fixing the relative phase $\Delta\theta$ between the two $\pi/2$ pulses. The visibility of the interference after a storage time of 5, 30 and 60 minutes is $V=93.0\pm5.5\%$, $95.3\pm4.2\%$ and $92.9\pm4.9\%$, corresponding to a fidelity of $F=(1+V)/2=96.5\pm2.8\%$, $97.6\pm2.1\%$ and $96.4\pm2.5\%$, respectively, which indicates a promising application as a faithful quantum memory for time-bin qubits in the future. Compared with the longest optical storage time demonstrated so far, which is one minute realized in Pr$^{3+}$:Y$_2$SiO$_5$ \cite{longlifeqm}, this coherent storage time is approximately enhanced by 60 times.

\section*{Discussion}

In summary, spin-wave AFC protocol is combined with ZEFOZ technique for the purpose of long-lived quantum storage of photonic qubits. An optical memory with a coherent optical storage time of one hour and a time-bandwidth product of $3.6\times10^9$ is achieved using a $^{151}$Eu$^{3+}$:Y$_2$SiO$_5$ crystal. In order to extend this work to single-photon regime, higher efficiencies and good signal-to-noise ratios (SNR) are required. There are two sources of noise when it comes to single-photon regime. One is the coherent noise caused by the intense control pulse, which can be filtered out with spectral holes \cite{pryso-sw-qm}, Fabry-P\'erot cavities \cite{euyso151-swafc-dd}, etc. The other is the photon noise caused by extra population in the ground state introduced by the DD sequences \cite{dd_noise}, which can be suppressed by better optical pumping and high-fidelity $\pi$ pulses. The storage time in this work can be further extended by using a stronger ZEFOZ field \cite{euyso-ground_rhs}, where a higher SNR can be expected since less DD pulses are required.

\textbf{Data availability} The data that support the findings of this study are available from the corresponding author upon reasonable request.

\bibliography{bibliography}

\begin{thebibliography}{10}
\expandafter\ifx\csname url\endcsname\relax
  \def\url#1{\texttt{#1}}\fi
\expandafter\ifx\csname urlprefix\endcsname\relax\def\urlprefix{URL }\fi
\providecommand{\bibinfo}[2]{#2}
\providecommand{\eprint}[2][]{\url{#2}}

\bibitem{repeater_1998}
\bibinfo{author}{Briegel, H.-J.}, \bibinfo{author}{Dür, W.},
  \bibinfo{author}{Cirac, J.~I.} \& \bibinfo{author}{Zoller, P.}
\newblock \bibinfo{title}{Quantum repeaters: The role of imperfect local
  operations in quantum communication}.
\newblock \emph{\bibinfo{journal}{Phys. Rev. Lett.}}
  \textbf{\bibinfo{volume}{81}}, \bibinfo{pages}{5932--5935}
  (\bibinfo{year}{1998}).

\bibitem{review_repeater}
\bibinfo{author}{Sangouard, N.}, \bibinfo{author}{Simon, C.},
  \bibinfo{author}{Riedmatten, H.~d.} \& \bibinfo{author}{Gisin, N.}
\newblock \bibinfo{title}{Quantum repeaters based on atomic ensembles and
  linear optics}.
\newblock \emph{\bibinfo{journal}{Rev. Mod. Phys.}}
  \textbf{\bibinfo{volume}{83}}, \bibinfo{pages}{33--80}
  (\bibinfo{year}{2011}).

\bibitem{network_duan}
\bibinfo{author}{Duan, L.-M.} \& \bibinfo{author}{Monroe, C.}
\newblock \bibinfo{title}{Colloquium: {Quantum} networks with trapped ions}.
\newblock \emph{\bibinfo{journal}{Rev. Mod. Phys.}}
  \textbf{\bibinfo{volume}{82}}, \bibinfo{pages}{1209--1224}
  (\bibinfo{year}{2010}).

\bibitem{single-atom_qm}
\bibinfo{author}{Specht, H.~P.} \emph{et~al.}
\newblock \bibinfo{title}{A single-atom quantum memory}.
\newblock \emph{\bibinfo{journal}{Nature}} \textbf{\bibinfo{volume}{473}},
  \bibinfo{pages}{190--193} (\bibinfo{year}{2011}).

\bibitem{entanglement_nv}
\bibinfo{author}{Kalb, N.} \emph{et~al.}
\newblock \bibinfo{title}{Entanglement distillation between solid-state quantum
  network nodes}.
\newblock \emph{\bibinfo{journal}{Science}} \textbf{\bibinfo{volume}{356}},
  \bibinfo{pages}{928--932} (\bibinfo{year}{2017}).

\bibitem{memory_entanglement}
\bibinfo{author}{Yu, Y.} \emph{et~al.}
\newblock \bibinfo{title}{Entanglement of two quantum memories via fibres over
  dozens of kilometres}.
\newblock \emph{\bibinfo{journal}{Nature}} \textbf{\bibinfo{volume}{578}},
  \bibinfo{pages}{240--245} (\bibinfo{year}{2020}).

\bibitem{space-borne}
\bibinfo{author}{Gündoğan, M.} \emph{et~al.}
\newblock \bibinfo{title}{Space-borne quantum memories for global quantum
  communication.} \bibinfo{note}{Preprint at http://arxiv.org/abs/2006.10636
  (2020)}.

\bibitem{satellite_entang}
\bibinfo{author}{Yin, J.} \emph{et~al.}
\newblock \bibinfo{title}{Satellite-based entanglement distribution over 1200
  kilometers}.
\newblock \emph{\bibinfo{journal}{Science}} \textbf{\bibinfo{volume}{356}},
  \bibinfo{pages}{1140--1144} (\bibinfo{year}{2017}).

\bibitem{6hour}
\bibinfo{author}{Zhong, M.} \emph{et~al.}
\newblock \bibinfo{title}{Optically addressable nuclear spins in a solid with a
  six-hour coherence time}.
\newblock \emph{\bibinfo{journal}{Nature}} \textbf{\bibinfo{volume}{517}},
  \bibinfo{pages}{177--180} (\bibinfo{year}{2015}).

\bibitem{longlifeqm}
\bibinfo{author}{Heinze, G.}, \bibinfo{author}{Hubrich, C.} \&
  \bibinfo{author}{Halfmann, T.}
\newblock \bibinfo{title}{Stopped light and image storage by
  electromagnetically induced transparency up to the regime of one minute}.
\newblock \emph{\bibinfo{journal}{Phys. Rev. Lett.}}
  \textbf{\bibinfo{volume}{111}}, \bibinfo{pages}{033601}
  (\bibinfo{year}{2013}).

\bibitem{my_2018}
\bibinfo{author}{Ma, Y.} \emph{et~al.}
\newblock \bibinfo{title}{A {Raman} heterodyne study of the hyperfine
  interaction of the optically-excited state $^5${D}$_0$ of
  {Eu}$^{3+}$:{Y}$_2${SiO}$_5$}.
\newblock \emph{\bibinfo{journal}{J. Lumin.}} \textbf{\bibinfo{volume}{202}},
  \bibinfo{pages}{32--37} (\bibinfo{year}{2018}).

\bibitem{euyso151-swafc-dd}
\bibinfo{author}{Jobez, P.} \emph{et~al.}
\newblock \bibinfo{title}{Coherent spin control at the quantum level in an
  ensemble-based optical memory}.
\newblock \emph{\bibinfo{journal}{Phys. Rev. Lett.}}
  \textbf{\bibinfo{volume}{114}}, \bibinfo{pages}{230502}
  (\bibinfo{year}{2015}).

\bibitem{rb87-1min}
\bibinfo{author}{Dudin, Y.~O.}, \bibinfo{author}{Li, L.} \&
  \bibinfo{author}{Kuzmich, A.}
\newblock \bibinfo{title}{Light storage on the time scale of a minute}.
\newblock \emph{\bibinfo{journal}{Phys. Rev. A}} \textbf{\bibinfo{volume}{87}},
  \bibinfo{pages}{031801} (\bibinfo{year}{2013}).

\bibitem{Rb_0.22s}
\bibinfo{author}{Yang, S.-J.}, \bibinfo{author}{Wang, X.-J.},
  \bibinfo{author}{Bao, X.-H.} \& \bibinfo{author}{Pan, J.-W.}
\newblock \bibinfo{title}{An efficient quantum light–matter interface with
  sub-second lifetime}.
\newblock \emph{\bibinfo{journal}{Nat. Photon.}} \textbf{\bibinfo{volume}{10}},
  \bibinfo{pages}{381--384} (\bibinfo{year}{2016}).

\bibitem{afcexp_1st}
\bibinfo{author}{de~Riedmatten, H.}, \bibinfo{author}{Afzelius, M.},
  \bibinfo{author}{Staudt, M.~U.}, \bibinfo{author}{Simon, C.} \&
  \bibinfo{author}{Gisin, N.}
\newblock \bibinfo{title}{A solid-state light-matter interface at the
  single-photon level}.
\newblock \emph{\bibinfo{journal}{Nature}} \textbf{\bibinfo{volume}{456}},
  \bibinfo{pages}{773--777} (\bibinfo{year}{2008}).

\bibitem{afc_theo}
\bibinfo{author}{Afzelius, M.}, \bibinfo{author}{Simon, C.},
  \bibinfo{author}{de~Riedmatten, H.} \& \bibinfo{author}{Gisin, N.}
\newblock \bibinfo{title}{Multimode quantum memory based on atomic frequency
  combs}.
\newblock \emph{\bibinfo{journal}{Phys. Rev. A}} \textbf{\bibinfo{volume}{79}},
  \bibinfo{pages}{052329} (\bibinfo{year}{2009}).

\bibitem{pryso-sw-qm}
\bibinfo{author}{Gundogan, M.}, \bibinfo{author}{Ledingham, P.~M.},
  \bibinfo{author}{Kutluer, K.}, \bibinfo{author}{Mazzera, M.} \&
  \bibinfo{author}{de~Riedmatten, H.}
\newblock \bibinfo{title}{Solid state spin-wave quantum memory for time-bin
  qubits}.
\newblock \emph{\bibinfo{journal}{Phys. Rev. Lett.}}
  \textbf{\bibinfo{volume}{114}}, \bibinfo{pages}{230501}
  (\bibinfo{year}{2015}).

\bibitem{time_entang}
\bibinfo{author}{Kutluer, K.} \emph{et~al.}
\newblock \bibinfo{title}{Time entanglement between a photon and a spin wave in
  a multimode solid-state quantum memory}.
\newblock \emph{\bibinfo{journal}{Phys. Rev. Lett.}}
  \textbf{\bibinfo{volume}{123}}, \bibinfo{pages}{030501}
  (\bibinfo{year}{2019}).

\bibitem{multiplexed_yang}
\bibinfo{author}{Yang, T.-S.} \emph{et~al.}
\newblock \bibinfo{title}{Multiplexed storage and real-time manipulation based
  on a multiple degree-of-freedom quantum memory}.
\newblock \emph{\bibinfo{journal}{Nat. Commun.}} \textbf{\bibinfo{volume}{9}},
  \bibinfo{pages}{3407} (\bibinfo{year}{2018}).

\bibitem{EuYSO_multiplexed}
\bibinfo{author}{Laplane, C.} \emph{et~al.}
\newblock \bibinfo{title}{Multiplexed on-demand storage of polarization qubits
  in a crystal}.
\newblock \emph{\bibinfo{journal}{New J. Phys.}} \textbf{\bibinfo{volume}{18}},
  \bibinfo{pages}{013006} (\bibinfo{year}{2015}).

\bibitem{euyso-correlation}
\bibinfo{author}{Laplane, C.}, \bibinfo{author}{Jobez, P.},
  \bibinfo{author}{Etesse, J.}, \bibinfo{author}{Gisin, N.} \&
  \bibinfo{author}{Afzelius, M.}
\newblock \bibinfo{title}{Multimode and long-lived quantum correlations between
  photons and spins in a crystal}.
\newblock \emph{\bibinfo{journal}{Phys. Rev. Lett.}}
  \textbf{\bibinfo{volume}{118}}, \bibinfo{pages}{210501}
  (\bibinfo{year}{2017}).

\bibitem{euyso-0.53s}
\bibinfo{author}{Holzäpfel, A.} \emph{et~al.}
\newblock \bibinfo{title}{Optical storage for 0.53 s in a solid-state atomic
  frequency comb memory using dynamical decoupling}.
\newblock \emph{\bibinfo{journal}{New J. Phys.}} \textbf{\bibinfo{volume}{22}},
  \bibinfo{pages}{063009} (\bibinfo{year}{2020}).

\bibitem{euyso-ground_rhs}
\bibinfo{author}{Longdell, J.~J.}, \bibinfo{author}{Alexander, A.~L.} \&
  \bibinfo{author}{Sellars, M.~J.}
\newblock \bibinfo{title}{Characterization of the hyperfine interaction in
  europium-doped yttrium orthosilicate and europium chloride hexahydrate}.
\newblock \emph{\bibinfo{journal}{Phys. Rev. B}} \textbf{\bibinfo{volume}{74}},
  \bibinfo{pages}{195101} (\bibinfo{year}{2006}).

\bibitem{euysoexcitedgisin}
\bibinfo{author}{Cruzeiro, E.~Z.} \emph{et~al.}
\newblock \bibinfo{title}{Characterization of the hyperfine interaction of the
  excited $^5${D}$_0$ state of {Eu}$^{3+}$:{Y}$_2${SiO}$_5$}.
\newblock \emph{\bibinfo{journal}{Phys. Rev. B}} \textbf{\bibinfo{volume}{97}},
  \bibinfo{pages}{094416} (\bibinfo{year}{2018}).

\bibitem{ramanprl}
\bibinfo{author}{Mlynek, J.}, \bibinfo{author}{Wong, N.~C.},
  \bibinfo{author}{DeVoe, R.~G.}, \bibinfo{author}{Kintzer, E.~S.} \&
  \bibinfo{author}{Brewer, R.~G.}
\newblock \bibinfo{title}{Raman heterodyne detection of nuclear magnetic
  resonance}.
\newblock \emph{\bibinfo{journal}{Phys. Rev. Lett.}}
  \textbf{\bibinfo{volume}{50}}, \bibinfo{pages}{993--996}
  (\bibinfo{year}{1983}).

\bibitem{SHBEuYSO}
\bibinfo{author}{Lauritzen, B.} \emph{et~al.}
\newblock \bibinfo{title}{Spectroscopic investigations of
  {Eu}$^{3+}$:{Y}$_2${SiO}$_5$ for quantum memory applications}.
\newblock \emph{\bibinfo{journal}{Phys. Rev. B}} \textbf{\bibinfo{volume}{85}},
  \bibinfo{pages}{115111} (\bibinfo{year}{2012}).

\bibitem{CHS}
\bibinfo{author}{Rippe, L.}, \bibinfo{author}{Nilsson, M.},
  \bibinfo{author}{Kröll, S.}, \bibinfo{author}{Klieber, R.} \&
  \bibinfo{author}{Suter, D.}
\newblock \bibinfo{title}{Experimental demonstration of efficient and selective
  population transfer and qubit distillation in a rare-earth-metal-ion-doped
  crystal}.
\newblock \emph{\bibinfo{journal}{Phys. Rev. A}} \textbf{\bibinfo{volume}{71}},
  \bibinfo{pages}{062328} (\bibinfo{year}{2005}).

\bibitem{euyso_spectra-t1-23d}
\bibinfo{author}{Könz, F.} \emph{et~al.}
\newblock \bibinfo{title}{Temperature and concentration dependence of optical
  dephasing, spectral-hole lifetime, and anisotropic absorption in
  {Eu}$^{3+}$:{Y}$_2${SiO}$_5$}.
\newblock \emph{\bibinfo{journal}{Phys. Rev. B}} \textbf{\bibinfo{volume}{68}},
  \bibinfo{pages}{085109} (\bibinfo{year}{2003}).

\bibitem{pryso_cvty}
\bibinfo{author}{Sabooni, M.}, \bibinfo{author}{Li, Q.},
  \bibinfo{author}{Kroll, S.} \& \bibinfo{author}{Rippe, L.}
\newblock \bibinfo{title}{Efficient quantum memory using a weakly absorbing
  sample}.
\newblock \emph{\bibinfo{journal}{Phys. Rev. Lett.}}
  \textbf{\bibinfo{volume}{110}}, \bibinfo{pages}{133604}
  (\bibinfo{year}{2013}).

\bibitem{dd_noise}
\bibinfo{author}{Cruzeiro, E.~Z.}, \bibinfo{author}{Fröwis, F.},
  \bibinfo{author}{Timoney, N.} \& \bibinfo{author}{Afzelius, M.}
\newblock \bibinfo{title}{Noise in optical quantum memories based on dynamical
  decoupling of spin states}.
\newblock \emph{\bibinfo{journal}{J. Mod. Optic.}}
  \textbf{\bibinfo{volume}{63}}, \bibinfo{pages}{2101--2113}
  (\bibinfo{year}{2016}).

\end{thebibliography}


\begin{thebibliography}{10}
\expandafter\ifx\csname url\endcsname\relax
  \def\url#1{\texttt{#1}}\fi
\expandafter\ifx\csname urlprefix\endcsname\relax\def\urlprefix{URL }\fi
\providecommand{\bibinfo}[2]{#2}
\providecommand{\eprint}[2][]{\url{#2}}

\bibitem{my_2018}
\bibinfo{author}{Ma, Y.} \emph{et~al.}
\newblock \bibinfo{title}{A {Raman} heterodyne study of the hyperfine
  interaction of the optically-excited state $^5${D}$_0$ of
  {Eu}$^{3+}$:{Y}$_2${SiO}$_5$}.
\newblock \emph{\bibinfo{journal}{J. Lumin.}} \textbf{\bibinfo{volume}{202}},
  \bibinfo{pages}{32--37} (\bibinfo{year}{2018}).

\bibitem{ramanprl}
\bibinfo{author}{Mlynek, J.}, \bibinfo{author}{Wong, N.~C.},
  \bibinfo{author}{DeVoe, R.~G.}, \bibinfo{author}{Kintzer, E.~S.} \&
  \bibinfo{author}{Brewer, R.~G.}
\newblock \bibinfo{title}{Raman heterodyne detection of nuclear magnetic
  resonance}.
\newblock \emph{\bibinfo{journal}{Phys. Rev. Lett.}}
  \textbf{\bibinfo{volume}{50}}, \bibinfo{pages}{993--996}
  (\bibinfo{year}{1983}).

\bibitem{euysoexcitedgisin}
\bibinfo{author}{Cruzeiro, E.~Z.} \emph{et~al.}
\newblock \bibinfo{title}{Characterization of the hyperfine interaction of the
  excited $^5${D}$_0$ state of {Eu}$^{3+}$:{Y}$_2${SiO}$_5$}.
\newblock \emph{\bibinfo{journal}{Phys. Rev. B}} \textbf{\bibinfo{volume}{97}},
  \bibinfo{pages}{094416} (\bibinfo{year}{2018}).

\bibitem{6hour}
\bibinfo{author}{Zhong, M.} \emph{et~al.}
\newblock \bibinfo{title}{Optically addressable nuclear spins in a solid with a
  six-hour coherence time}.
\newblock \emph{\bibinfo{journal}{Nature}} \textbf{\bibinfo{volume}{517}},
  \bibinfo{pages}{177--180} (\bibinfo{year}{2015}).

\bibitem{SHBEuYSO}
\bibinfo{author}{Lauritzen, B.} \emph{et~al.}
\newblock \bibinfo{title}{Spectroscopic investigations of
  {Eu}$^{3+}$:{Y}$_2${SiO}$_5$ for quantum memory applications}.
\newblock \emph{\bibinfo{journal}{Phys. Rev. B}} \textbf{\bibinfo{volume}{85}},
  \bibinfo{pages}{115111} (\bibinfo{year}{2012}).

\bibitem{afc_parallel}
\bibinfo{author}{Jobez, P.} \emph{et~al.}
\newblock \bibinfo{title}{Towards highly multimode optical quantum memory for
  quantum repeaters}.
\newblock \emph{\bibinfo{journal}{Phys. Rev. A}} \textbf{\bibinfo{volume}{93}},
  \bibinfo{pages}{032327} (\bibinfo{year}{2016}).

\bibitem{afc_theo}
\bibinfo{author}{Afzelius, M.}, \bibinfo{author}{Simon, C.},
  \bibinfo{author}{de~Riedmatten, H.} \& \bibinfo{author}{Gisin, N.}
\newblock \bibinfo{title}{Multimode quantum memory based on atomic frequency
  combs}.
\newblock \emph{\bibinfo{journal}{Phys. Rev. A}} \textbf{\bibinfo{volume}{79}},
  \bibinfo{pages}{052329} (\bibinfo{year}{2009}).

\bibitem{CHS}
\bibinfo{author}{Rippe, L.}, \bibinfo{author}{Nilsson, M.},
  \bibinfo{author}{Kröll, S.}, \bibinfo{author}{Klieber, R.} \&
  \bibinfo{author}{Suter, D.}
\newblock \bibinfo{title}{Experimental demonstration of efficient and selective
  population transfer and qubit distillation in a rare-earth-metal-ion-doped
  crystal}.
\newblock \emph{\bibinfo{journal}{Phys. Rev. A}} \textbf{\bibinfo{volume}{71}},
  \bibinfo{pages}{062328} (\bibinfo{year}{2005}).

\bibitem{nutation_spin_echo}
\bibinfo{author}{Ardelean, I.}, \bibinfo{author}{Kimmich, R.} \&
  \bibinfo{author}{Klemm, A.}
\newblock \bibinfo{title}{The nutation spin echo and its use for localized
  nmr}.
\newblock \emph{\bibinfo{journal}{J. Magn. Reson.}}
  \textbf{\bibinfo{volume}{146}}, \bibinfo{pages}{43--48}
  (\bibinfo{year}{2000}).

\bibitem{phase_memory}
\bibinfo{author}{Mims, W.~B.}
\newblock \bibinfo{title}{Phase memory in electron spin echoes, lattice
  relaxation effects in {CaWO}$_4$:{Er}, {Ce}, {Mn}}.
\newblock \emph{\bibinfo{journal}{Phys. Rev.}} \textbf{\bibinfo{volume}{168}},
  \bibinfo{pages}{370--389} (\bibinfo{year}{1968}).

\end{thebibliography}

\begin{acknowledgments}

\textbf{Acknowledgements} This work is supported by the National Key R\&D Program of China (No.~2017YFA0304100), the National Natural Science Foundation of China (Nos.~11774331, 11774335, 11504362, 11821404 and 11654002), Anhui Initiative in Quantum Information Technologies (No.~AHY020100), the Key Research Program of Frontier Sciences, CAS (No.~QYZDY-SSW-SLH003), the Science Foundation of the CAS (No.~ZDRW-XH-2019-1)  and the Fundamental Research Funds for the Central Universities (No.~WK2470000026, No.~WK2470000029). Z.-Q.Z acknowledges the support from the Youth Innovation Promotion Association CAS.

\textbf{Author contributions} Z.-Q.Z. and C.-F.L. conceived the experiment. Y.M. and Y.-Z.M. performed the experiment. Y.M. and Z.-Q.Z. wrote the manuscript. Z.-Q.Z., C.-F.L. and G.-C.G. supervised the project. All authors discussed the experimental procedures and results. 

\textbf{Competing interests} The authors declare no competing interests. 

\end{acknowledgments}

\end{document}

% \preprint{APS/123-QED}

\title{Supplementary Information for \\
One-hour coherent optical storage in an atomic frequency comb memory}% Force line breaks with \\

\author{Yu Ma}
\author{You-Zhi Ma}
\author{Zong-Quan Zhou}
 \email{zq\_zhou@ustc.edu.cn}
\author{Chuan-Feng Li}
 \email{cfli@ustc.edu.cn}
\author{Guang-Can Guo}
\affiliation{
 CAS Key Laboratory of Quantum Information,\\University of Science and Technology of China, Hefei, 230026, P. R. China
}
\affiliation{
 CAS Center For Excellence in Quantum Information and Quantum Physics, University of Science and Technology of China, Hefei, 230026, P. R. China
}

% \author{Ann Author}
%  \altaffiliation[Also at ]{Physics Department, XYZ University.}%Lines break automatically or can be forced with \\
%  \email{Second.Author@institution.edu}
% \affiliation{%
%  Authors' institution and/or address\\
%  This line break forced with \textbackslash\textbackslash
% }%

% \collaboration{MUSO Collaboration}%\noaffiliation

%  \homepage{http://www.Second.institution.edu/~Charlie.Author}
% \affiliation{
%  Second institution and/or address\\
%  This line break forced% with \\
% }%
% \affiliation{
%  Third institution, the second for Charlie Author
% \affiliation{%
%  Authors' institution and/or address\\
%  This line break forced with \textbackslash\textbackslash
% }%

% \collaboration{CLEO Collaboration}%\noaffiliation

%              %  but any date may be explicitly specified

% \begin{abstract}
% An article usually includes an abstract, a concise summary of the work
% covered at length in the main body of the article. 
% \begin{description}
% Secondary publications and information retrieval purposes.
% You may use the \texttt{description} environment to structure your abstract;
% use the optional argument of the \verb+\item+ command to give the category of each item. 
% \end{description}
% \end{abstract}

\maketitle

%\tableofcontents

\section*{\label{sec:prhd}S\lowercase{upplementary }N\lowercase{ote 1 - }\\P\lowercase{ulsed} R\lowercase{aman heterodyne detection}}

In our previous work \cite{my_2018}, we use Raman heterodyne detection method to probe the continuous-wave nuclear magnetic resonance (NMR) of the ground state $^7$F$_0$ and excited state $^5$D$_0$ of $^{151}$Eu$^{3+}$:Y$_2$SiO$_5$. However, in this work, although the ground-state resonances can be detected using the same method, the excited-state resonances are difficult to detect due to the weaker radio-frequency (RF) fields generated by the Helmholtz coils. 
% In the previous work, we use a solenoid coil wrapping around the sample, which has a stronger interaction with the doped ions.
% this method is difficult to detect the excited-state resonances due to the short-lived population in the optically excited states and the weaker interaction strength with the radio-frequency (RF) field.

%  In order to increase the signal to noise ratio (SNR), we first use 
To overcome this problem, we use a pulsed Raman heterodyne detection approach \cite{ramanprl}. Optical pumping is employed to isolate a single class of ions in a specific ground-state level via class cleaning and spin polarization described in the paper. Then a probe pulse with a duration of 200 $\mu$s and a power of 4 mW resonant with the populated ground state and an arbitrary excited state is sent into the crystal, and at the mean time an RF pulse with a duration of 100 $\mu$s is applied to drive the hyperfine transitions in the excited state. The RF pulse is generated by a local oscillator with an amplitude of 0.6 V, and amplified by an amplifier with an output of 300 W to ensure a strong enough interaction with the excited-state hyperfine levels. The transmitted optical pulse is accompanied with a Raman scattered field if the RF pulse is resonant with the hyperfine transitions. A beat signal between the transmitted and scattered field is then captured by a photodetector and demodulated with a lock-in amplifier.

% \begin{figure*}[hbt]
% \includegraphics[width=\textwidth]{SM-fig1v1.pdf}
% \caption{\label{fig:prhd}Raman heterodyne detection of pulsed NMR of the $\ket{3}_e\leftrightarrow\ket{6}_e$ transition.}
% \end{figure*}

% Note that the RF pulse we use here is single-frequency with a bandwidth of $1/100=0.01$ MHz. 
%  of a 124.52 MHz signal
The resonance of the hyperfine transitions can be identified by varying the frequency of RF pulses. At least five resonances need to be found in order to determine the excited-state structure with six hyperfine levels. An example of the $\ket{3}_\mathrm{e}\leftrightarrow\ket{6}_\mathrm{e}$ transition at 124.52 MHz is shown in Supplementary Fig.~{\ref{fig:prhd}}. All the neighboring hyperfine transitions in the excited state are presented in Supplementary Table~\ref{tab:fitable}. As a comparison, we also provide the calculated results in the same field, based on the spin Hamiltonians determined in previous works \cite{euysoexcitedgisin,my_2018}. Note that due to the experimental errors, the calculated strength and direction of the ZEFOZ field is slightly different based on these two works. But we verify that the first-order Zeeman coefficient $S_1$ is the same, which ensures that the fields are physically equivalent. According to the data presented in Supplementary Table~\ref{tab:fitable}, the maximum error is larger than 1 MHz, which is the storage bandwidth of the memory. In addition, we note that the level structure of the $^7$F$_0$ ground state obtained here is significantly different from the results that reported in Ref.~\cite{6hour}. These results indicate the necessity of this experiment before determining the optical pumping strategies.
% The experimentally determined results in the ZEFOZ magnetic field used in the paper
% In our previous work \cite{my_2018}, the predicted frequency of $\ket{3}_e\leftrightarrow\ket{6}_e$ is $23.86+79.86+20.87=124.59$ MHz, which is close to the pulsed NMR experimental result. 
% With a good SNR, we verify that the energy level structure is consistent with the theoretical predictions in the previous work shown in TABLE \ref{fitable}, with an error less than 200 kHz.

\begin{table}[tb]
\renewcommand{\arraystretch}{1.2}
\caption{Experimental results of the transition frequency between the neighboring energy levels in the ZEFOZ field, as a comparison with the calculated results based on the fitted Hamiltonians determined in the previous work (calc. I and II refer to calculation based on [\onlinecite{euysoexcitedgisin}] and [\onlinecite{my_2018}], respectively). Unit: MHz.}\label{tab:fitable}
\centering
\begin{tabular}{SSSS} \toprule
    %  & \multicolumn{2}{c}{Excited state $^5$D$_0$}      \\
    {Transition}        &  {Freq. (exp.)}    & {Freq. (calc. I)}    & {Freq. (calc. II)} \\\midrule
    {$\ket{5}_\mathrm{e}\leftrightarrow\ket{6}_\mathrm{e}$}       & 20.725         & 21.908      & 20.865    \\
    {$\ket{4}_\mathrm{e}\leftrightarrow\ket{5}_\mathrm{e}$}       & 79.903         & 79.899      & 79.856    \\
    {$\ket{3}_\mathrm{e}\leftrightarrow\ket{4}_\mathrm{e}$}       & 23.887         & 23.268      & 23.858    \\
    {$\ket{2}_\mathrm{e}\leftrightarrow\ket{3}_\mathrm{e}$}       & 56.199         & 55.999      & 56.089    \\
    {$\ket{1}_\mathrm{e}\leftrightarrow\ket{2}_\mathrm{e}$}       & 23.776         & 23.860      & 23.939    \\\bottomrule
\end{tabular}
\end{table}

\begin{figure*}[htb]
\includegraphics[width=\textwidth]{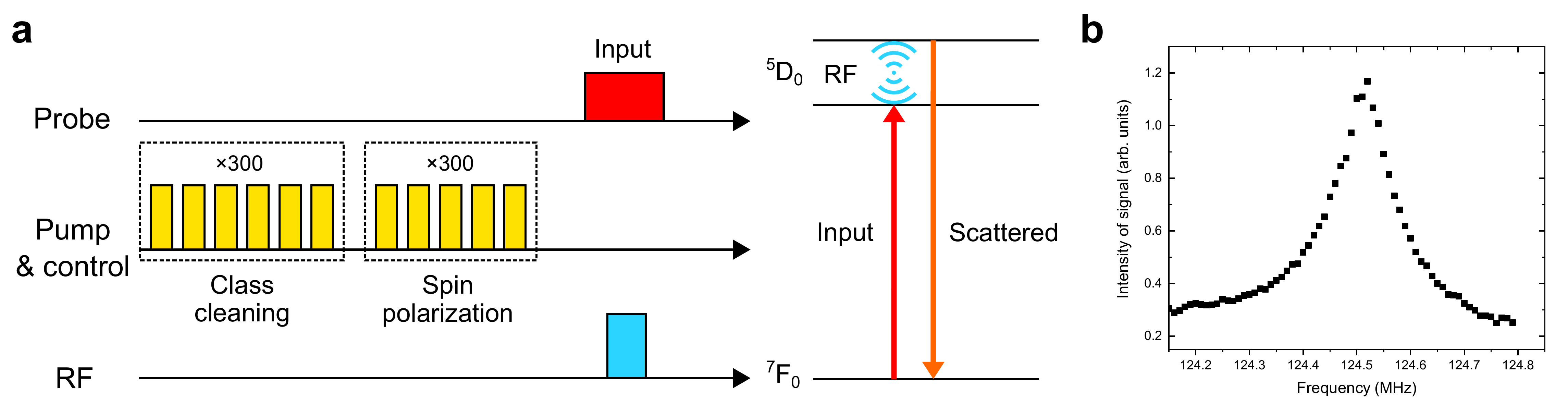}
\caption{\label{fig:prhd}\textbf{Experimental sequence and results of the pulsed Raman heterodyne detection.} \textbf{a} The experimental sequence of the pulsed Raman heterodyne detection of the excited-state NMR in the ZEFOZ magnetic field. Class cleaning and spin polarization are employed to isolate a single class of ions in a specific ground-state energy level. The input pulse excites the ions to the excited state while an RF pulse simultaneously drives the excited-state hyperfine transitions. The scattered field is generated when the RF pulse is resonant with a specific excited-state transition. \textbf{b} Raman heterodyne detection of pulsed NMR of the $\ket{3}_\mathrm{e}\leftrightarrow\ket{6}_\mathrm{e}$ transition.}
% \includegraphics[width=0.5\textwidth]{SM-fig2v1.pdf}
% \caption{\label{fig:afc}(Color online) The optical depth of the $\ket{4}_g$ (green) and $\ket{3}_g$ (blue), respectively, after class cleaning and spin polarization on $\ket{3}_g$. The $\ket{4}_g$ is cleaned out for the following spin wave storage. An AFC structure with ten teeth on $\ket{3}_g$ is shown by the red line.}
% (b) The spin-wave AFC echo as a function of the spin storage time. The fitted spin inhomogeneous broadening is 25.3 kHz. Some points are omitted around 10 $\mu$s because they are overlapped with the second-order two-level echo, which is emitted at 20 $\mu$s after the input pulse. Each point is averaged eight times.
% \includegraphics[width=0.5\textwidth]{SM-fig3v1.pdf}
% \caption{\label{fig:spinecho}(Color online) Decay of the spin echo with different dynamical decoupling sequences. CPMG (gray) has the longest coherence time of $3.22\pm0.11$ hours. KDD$_x$ (red) and XY4 (blue) have a coherence time of $0.92\pm0.05$ and $1.64\pm0.12$ hours, respectively. Each point is averaged three times.}
\end{figure*}

\begin{figure*}[hbt]
\includegraphics[width=\textwidth]{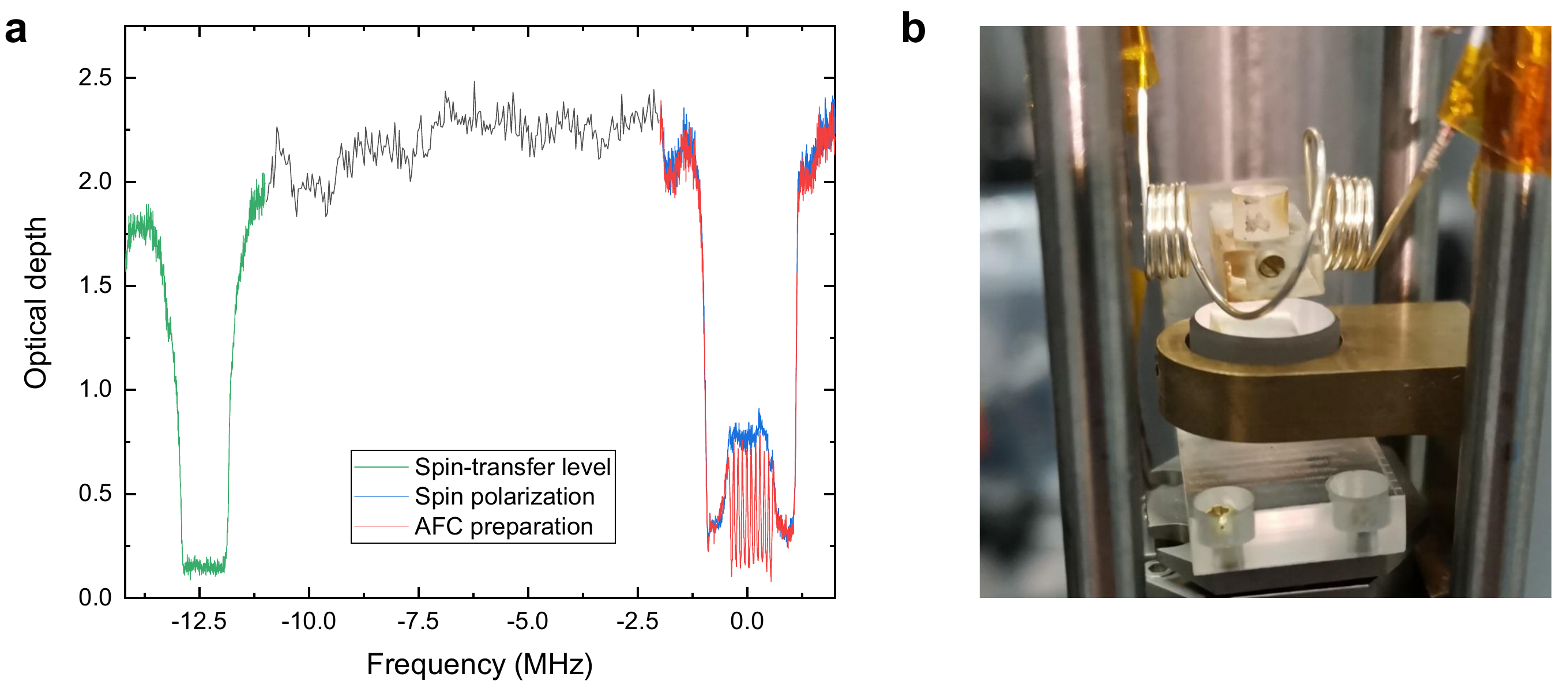}
\caption{\label{fig:afc}\textbf{AFC structure and a photo of the sample.} \textbf{a} The optical depth of the $\ket{4}_\mathrm{g}\leftrightarrow\ket{3}_\mathrm{e}$ (green) and $\ket{3}_\mathrm{g}\leftrightarrow\ket{3}_\mathrm{e}$ (blue) after the class cleaning and spin polarization, and the AFC structure with ten teeth prepared in the $\ket{3}_\mathrm{g}$ (red). \textbf{b} The photo of the sample, which is fixed on a specially-designed holder mounted on the goniometers. The helmholtz coils are placed at the two sides of the sample and the mirror is below the sample. The external magnetic field and the incident beams are along the vertical direction.}
% (b) The spin-wave AFC echo as a function of the spin storage time. The fitted spin inhomogeneous broadening is 25.3 kHz. Some points are omitted around 10 $\mu$s because they are overlapped with the second-order two-level echo, which is emitted at 20 $\mu$s after the input pulse. Each point is averaged eight times.
% \includegraphics[width=0.5\textwidth]{SM-fig3v1.pdf}
% \caption{\label{fig:spinecho}(Color online) Decay of the spin echo with different dynamical decoupling sequences. CPMG (gray) has the longest coherence time of $3.22\pm0.11$ hours. KDD$_x$ (red) and XY4 (blue) have a coherence time of $0.92\pm0.05$ and $1.64\pm0.12$ hours, respectively. Each point is averaged three times.}
\end{figure*}

\begin{figure*}[ht]
\includegraphics[width=\textwidth]{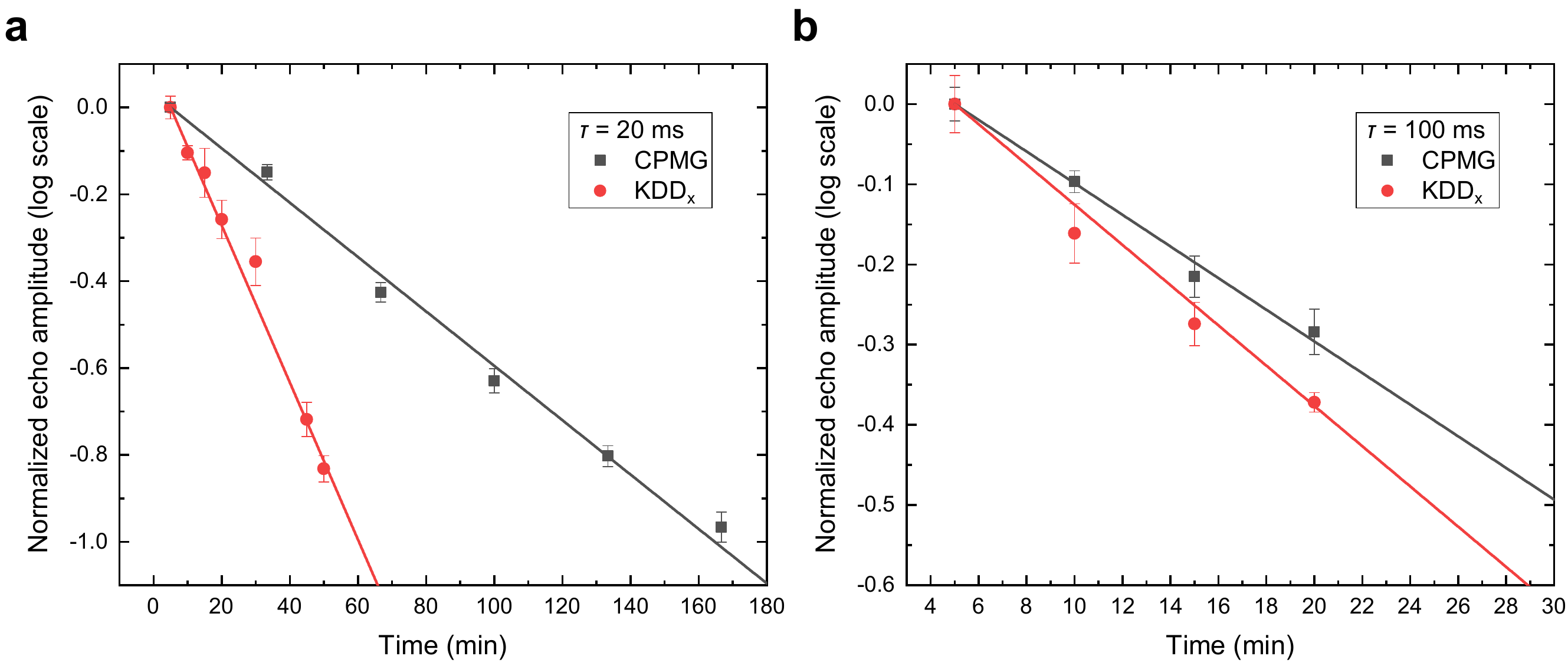}
\caption{\label{fig:spinecho}\textbf{Measurements of the spin coherence time in the ZEFOZ field.} \textbf{a} Decay of the spin echo with different dynamical decoupling sequences. CPMG (gray) has the longest coherence time of $2.68\pm0.06$ hours. KDD$_\mathrm{x}$ (red) has a coherence time of $0.95\pm0.03$. \textbf{b} When $\tau = 100$ ms, the coherence times are $50.6\pm2.0$ and $38.2\pm2.0$ minutes with CPMG and KDD$_\mathrm{x}$, respectively. Each data point is averaged three times, and the error bars indicate one standard deviation of the echo area.}
\end{figure*}

% \section{\label{sec:afc}Atomic frequency comb}
\section*{\label{sec:afc}S\lowercase{upplementary }N\lowercase{ote 2 - }\\A\lowercase{tomic frequency comb}}

The laser system at 580 nm is a specially-designed high-power laser. The seed laser is a semiconductor laser at 1160 nm (DL pro, Toptica) which is amplified to 30 W by a Raman fiber amplifier (PreciLasers). The laser at 580 nm is obtained by the single-pass second harmonic generation in a PPSLT crystal, with an output power of 6.2 W. The reference optical pulse for heterodyne detection is generated from another frequency-doubled semiconductor laser (TA-SHG, Toptica) with a locked frequency offset.
% The laser used to implement the AFC preparation and storage at 580 nm is frequency doubled from a semiconductor laser at 1160 nm (TOPTICA Photonics), which is amplified by a Raman amplifier (PreciLasers). The maximum output power of the laser is 1.6 W.

Before preparing the AFC in the $\ket{3}_\mathrm{g}$ level, optical pumping sequence called class cleaning and spin polarization \cite{SHBEuYSO} is used to select one class of ions. The six pumping lights are resonant with $\ket{1}_\mathrm{g}\leftrightarrow\ket{1}_\mathrm{e}$, $\ket{2}_\mathrm{g}\leftrightarrow\ket{2}_\mathrm{e}$, $\ket{3}_\mathrm{g}\leftrightarrow\ket{3}_\mathrm{e}$, $\ket{4}_\mathrm{g}\leftrightarrow\ket{3}_\mathrm{e}$, $\ket{5}_\mathrm{g}\leftrightarrow\ket{3}_\mathrm{e}$, $\ket{6}_\mathrm{g}\leftrightarrow\ket{5}_\mathrm{e}$, respectively, with a power of 100 $\mu$W and a duration of 2 ms.
% Each pump light is chirped over 1 MHz to create a bandwidth for the following AFC preparation. The bandwidth cannot be extended in this optical pumping scheme because a larger bandwidth of the chirp will be resonant with other classes of ions and reduce the transfer efficiency of the control pulses.

The structure of the AFC is presented in Supplementary Fig.~\ref{fig:afc}a. We use a single-photon detector to record the absorption profile of the AFC prepared in $\ket{3}_\mathrm{g}$. The AFC is prepared using a parallel method \cite{afc_parallel} within the 1 MHz bandwidth. The AFC has a comb periodicity of 100 kHz and a single peak width (FWHM) of $\gamma=45$ kHz, leading to a finesse $F=100/45=2.22$. The two-level AFC efficiency can be calculated as follows: \cite{afc_theo}
% \begin{equation}
%     \eta = (1-e^{-(\alpha L/F)(\sqrt{\pi/4 ln2})})^2 e^{-(1/F^2)(\pi^2/2 ln2)},
% \end{equation}
\begin{equation}
\eta=\left(1-e^{-(\alpha L / F)(\sqrt{\pi / 4 \ln 2})}\right)^{2} e^{-\left(1 / F^{2}\right)\left(\pi^{2} / 2 \ln 2\right)},
\end{equation}
where the optical depth after class cleaning and spin polarization is $\alpha L=0.8$. The calculated efficiency $\eta$ is 4.4\%, which is close to our experimental result of 4.5\%.
% The efficiency is estimated by calculating the ratio of the intensity of the echo to the intensity of an transmitted reference pulse. The intensity of the reference pulse is measured when the laser is tuned off-resonant to prevent the crystal from absorbing the reference pulse while keeping the same power as the input pulse incident into the AFC.

The control pulse used in the paper has a complex hyperbolic secant profile \cite{CHS} to achieve efficient control over a large bandwidth. The transfer efficiency $\eta_{\mathrm{control}}$ of a single control pulse is estimated by comparing the spin-wave AFC echo with the two-level AFC echo.

% The input and control mode are both linearly polarized. The polarization of the input mode is adjusted to reach a maximum absorption, and that of the control mode is approximately aligned with the input mode with a small angle of 4.2$^{\circ}$.
% Both the input and control mode are linearly polarized and approximately aligned with a small angle of 4.2$^{\circ}$.

In Supplementary Fig.~\ref{fig:afc}b, we present a photo of the cylindrical sample and its surroundings. The optical surface is cut perpendicularly to the direction of the known ZEFOZ magnetic field in order to easily align the sample to the magnetic field of the magnet, which is approximately in the vertical direction.

% \section{\label{sec:se}Spin coherence time measurements}
\section*{\label{sec:se}S\lowercase{upplementary }N\lowercase{ote 3 - }\\S\lowercase{pin coherence time measurements}}

The width of the $\pi$ pulse of 65.1 $\mu$s used in the dynamical decoupling and spin coherence time measurements is determined by implementing a spin nutation experiment \cite{nutation_spin_echo}. The two-pulse phase memory time $T_\mathrm{M}$ \cite{phase_memory} is determined to be 21.5 s, which is shorter than the 47 s in the previous work \cite{6hour}. The possible reason is the inhomogeneity of the magnetic field, considering the longer sample used here.
% This is because here we are using a thicker sample with a higher concentration, which results in a wider spin inhomogeneous broadening. In addition, the inhomogeneity of the field of both the magnet and the RF coils also reduces the phase memory time and coherence time of the thick sample.

In order to take advantage of the long spin coherence time of the $\ket{3}_\mathrm{g}\leftrightarrow\ket{4}_\mathrm{g}$ ($\ket{-3/2}_\mathrm{g}\leftrightarrow\ket{+3/2}_\mathrm{g}$) transition in the ZEFOZ field, we measure the coherence times by spin echo experiments with different dynamical decoupling sequences applied as shown in Supplementary Fig.~\ref{fig:spinecho}. The spin echo signals are detected by Raman heterodyne detection. CPMG and KDD$_\mathrm{x}$ reach their longest coherence times of $2.68\pm0.06$ and $0.95\pm0.03$ hours respectively, when the interval $\tau$ between the $\pi$ pulses is 20 ms. When $\tau$ is shorter than 20 ms, the heating of the Helmholtz coils and the infidelity of the $\pi$ pulses will limit the spin coherence times. When $\tau = 100$ ms, the coherence times are $50.6\pm2.0$ and $38.2\pm2.0$ minutes with CPMG and KDD$_\mathrm{x}$, respectively. The spin coherent lifetimes are essentially the same as the optical AFC storage lifetimes for $\tau=100$ ms. We note that, due to strong fluctuations of the signal for optical storage with storage time shorter than 5 minutes, the lifetimes are fitted using the data starting from 5 minutes in Fig.~3 in the main text.
% will heat up the sample and greatly shorten
% Among the three kinds of sequences, CPMG has the longest coherence time of $3.22\pm0.11$ hours.
% Here we use the same optically-detected method to probe the echo as in the previous work \cite{6hour}.

% This sample document demonstrates proper use of REV\TeX~4.2 (and
% \LaTeXe) in mansucripts prepared for submission to APS
% journals. Further information can be found in the REV\TeX~4.2
% documentation included in the distribution or available at
% \url{http://journals.aps.org/revtex/}.

% When commands are referred to in this example file, they are always
% shown with their required arguments, using normal \TeX{} format. In
% this format, \verb+#1+, \verb+#2+, etc. stand for required
% author-supplied arguments to commands. For example, in
% \verb+\section{#1}+ the \verb+#1+ stands for the title text of the
% author's section heading, and in \verb+\title{#1}+ the \verb+#1+
% stands for the title text of the paper.

% Line breaks in section headings at all levels can be introduced using
% \textbackslash\textbackslash. A blank input line tells \TeX\ that the
% paragraph has ended. Note that top-level section headings are
% automatically uppercased. If a specific letter or word should appear in
% lowercase instead, you must escape it using \verb+\lowercase{#1}+ as
% in the word ``via'' above.

% \nocite{*}

\bibliography{bibliography}% Produces the bibliography via BibTeX.